\documentclass{article}
\usepackage{authblk}
\usepackage{babel,spconf,amsmath,graphicx,hyperref,lipsum,enumitem,multicol,makecell}

\newcommand{\specialcell}[2][c]{%
  \begin{tabular}[#1]{@{}c@{}}#2\end{tabular}}

\usepackage{xcolor}
\usepackage{float}

\name{%
\begin{tabular}{@{}c@{}}
Ivan Yakovlev \qquad 
Mikhail Melnikov \qquad 
Nikita Bukhal \qquad 
Rostislav Makarov  \\ 
Alexander Alenin \qquad 
Nikita Torgashov \qquad 
Anton Okhotnikov
\end{tabular}}

\title{LRPD: Large Replay Parallel Dataset}
\address{ID R\&D Inc., New York, USA}

\begin{document}

\begin{titlepage}
    
\begin{center}
\Large \textbf{IEEE COPYRIGHT NOTICE}
\end{center}

\begin{enumerate}

\item Personal use of this material is permitted.  Permission from IEEE must be obtained for all other uses, in any current or future media, including reprinting/republishing this material for advertising or promotional purposes, creating new collective works, for resale or redistribution to servers or lists, or reuse of any copyrighted component of this work in other works. © 2023 IEEE.

\item This material is presented to ensure timely dissemination of scholarly and technical work. Copyright and all rights therein are retained by authors or by other copyright holders. All persons copying this information are expected to adhere to the terms and constraints invoked by each author’s copyright. In most cases, these works may not be reposted without the explicit permission of the copyright holder. For more details, see the IEEE Copyright Policy.

\item This work is presented and published at the proceedings of:
\begin{center}
\textbf{ICASSP 2022 - 2022 IEEE International Conference on Acoustics, Speech and Signal Processing (ICASSP)}
\end{center}

\item Link to the final version of the paper on IEEE Xplore: \url{https://ieeexplore.ieee.org/document/9746527}
\begin{enumerate}
\item Paper Title: \textbf{LRPD: Large Replay Parallel Dataset}

\item Authors: Ivan Yakovlev, Mikhail Melnikov, Nikita Bukhal, Rostislav Makarov, Alexander Alenin, Nikita Torgashov, Anton Okhotnikov

\item Paper ID: 4055

\item Acceptance Date: 21 January 2022

\item Conference Date: 22-27 May 2022

\item Conference Location: Singapore, Sands Expo \& Convention Centre

\item DOI: \href{https://doi.org/10.1109/ICASSP43922.2022.9746527}{10.1109/ICASSP43922.2022.9746527}

\item Publisher: IEEE

\item Cite as: 
\begin{verbatim}
Yakovlev, I., Melnikov, M., Bukhal, N., Makarov, R., Alenin, A., Torgashov, 
N., & Okhotnikov, A. (2022). LRPD: Large Replay Parallel Dataset. ICASSP 
2022 - 2022 IEEE International Conference on Acoustics, Speech and Signal 
Processing (ICASSP), 6612-6616. 
https://doi.org/10.1109/ICASSP43922.2022.9746527
\end{verbatim}

\item BibTeX:
\begin{verbatim}
@INPROCEEDINGS{9746527,
  author={Yakovlev, Ivan and Melnikov, Mikhail and Bukhal, Nikita and 
  Makarov, Rostislav and Alenin, Alexander and Torgashov, Nikita and 
  Okhotnikov, Anton},
  booktitle={ICASSP 2022 - 2022 IEEE International Conference on Acoustics, 
  Speech and Signal Processing (ICASSP)}, 
  title={LRPD: Large Replay Parallel Dataset}, 
  year={2022},
  volume={},
  number={},
  pages={6612-6616},
  doi={10.1109/ICASSP43922.2022.9746527}
}
\end{verbatim}

\end{enumerate}

\end{enumerate}

\end{titlepage}
%
\maketitle
\begin{abstract}




The latest research in the field of voice anti-spoofing (VAS) shows that deep neural networks (DNN) outperform classic approaches like GMM in the task of presentation attack detection. However, DNNs require a lot of data to converge, and still lack generalization ability. In order to foster the progress of neural network systems, we introduce a Large Replay Parallel Dataset (LRPD) aimed for a detection of replay attacks. LRPD contains more than 1M utterances collected by 19 recording devices in 17 various environments. We also provide an example training pipeline in PyTorch \cite{paszke2019pytorch} and a baseline system, that achieves 0.28\% Equal Error Rate (EER) on evaluation subset of LRPD and 11.91\% EER on publicly available ASVpoof 2017  \cite{kinnunen17_interspeech} eval set. These results show that model trained with LRPD dataset has a consistent performance on the fully unknown conditions. Our dataset is free for research purposes and hosted on GDrive\footnote[1]{\url{https://drive.google.com/drive/folders/1lHxQ5tPco5F1N8xv_x7lfDOhU2SEsOU9?usp=sharing}}. Baseline code and pre-trained models are available at GitHub\footnote[2]{\url{https://github.com/IDRnD/lrpd-paper-code}}. 


\end{abstract}
\section{Introduction}
\label{sec:intro}



Over the last years a lot of work has been done in the voice biometrics field. Recent speaker recognition systems are able to successfully recognize person using the voice over various conditions. Such progress allows to build  reliable voice-based solutions for person authentication. However, it is usually assumed that biometrics systems are vulnerable to spoofing attacks, also known as presentation attacks. For the best of our knowledge, there are two main types of spoofing attacks, logical and physical access attacks.

Logical access (LA) attach comprised two different approaches based on speech synthesis systems. Text-to-speech (TTS) systems are used to generate a fully artificial speech based on the specified text, while the voice conversion (VC) systems use a speech of one person as an input and convert it into the speech that resembles the voice of another person.

Physical access (PA) attacks are also known as Replay attacks, and are performed in the following way: the bona fide speech of the target speaker is recorded first, and then it is being presented to the speaker recognition system by a playback using the mobile phone or speaker. In this paper we focus on the problem of physical attacks, due to the ease of implementation and a high difficulty of detection.


There are multiple publicly available datasets for training presentation attack detection systems. The most widespread datasets are related to ASVspoof challenges, that made a huge contribution to the VAS research. The first challenge, ASVspoof 2015 \cite{wu2015asvspoof}, was focused on the Logical access attacks detection. The second challenge, ASVspoof 2017, was focused on the physical access attacks detection. The ASVspoof 2019 \cite{todisco2019asvspoof} challenge aimed to consider both types of presentation attacks. AVspoof \cite{ergunay2015vulnerability} is a public audio spoofing database which includes 10 various spoofing threats generated using replay, TTS and VC systems. VoicePA \cite{korshunov2018use} is an extension of the AVspoof database, which includes presentation attacks recorded in different environments with various recording and playback devices. PHONESPOOF \cite{lavrentyeva2019phonespoof} is a database that was collected in the telephone channel domain and is used to investigate the robustness of anti-spooﬁng systems in the telephone channel conditions.




The specific issue associated with replayed databases, is that they tend to become outdated over the time. The main difference between bona fide and replayed speech is a small distortion of source signal, caused by both recording and playback devices, which is likely to be used by anti-spoofing system to discriminate. Thus, even robust replay detection system may be vulnerable to the hardware or software speech preprocessing algorithms of new smartphones, which are being updated at least annually. We believe, that LRPD dataset, containing recent smartphones, with its wide covering of recording environments, is going to be very handful for the further development of replay detection systems.

In Section 2 of this paper we present the LRPD dataset. Section 3 describes experiments conducted on the LRPD and ASVspoof 2017 datasets. And finally, Sections 4 and 5 contain analysis of experiments' results and conclusions on our work.

\section{Dataset description}
The LRPD corpus was collected and open-sourced to push the boundaries of current research in the field of Replay spoofing attacks detection. The distinctive feature of LRPD dataset is that it contains several copies of replayed audio recorded by different devices at the same time in parallel.

\label{sec:description}

\subsection{Audio sources}

The LRPD dataset contains both bona fide and replayed types of utterances. Bona fide speech was taken from VCTK \cite{yamagishi2019vctk}, LibriSpeech \cite{panayotov2015librispeech}, Mozilla Common Voices (MCV) \cite{ardila2019common}, Google Language Resources (GLR) \cite{butryna2020google}, and CN-Celeb \cite{li2020cn} datasets. For each dataset we randomly sampled a subset of files to use. The resulting number of speakers and files selected for the replaying is shown in Table \ref{table:source-datasets}.


\begin{table}[t]
	\centering
	\caption{Source datasets}
	\label{table:source-datasets}
	\begin{tabular}{lcccc}
		\hline \hline
		& \multicolumn{2}{c}{\textbf{Train part}} & \multicolumn{2}{c}{\textbf{Eval part}} \\
		\textbf{Dataset name} & \specialcell{Spks} & \specialcell{Utts} & \specialcell{Spks} & \specialcell{Utts} \\ 
		\hline
        VCTK                  & -       & -               &  107   & 1172  \\
        LibriSpeech           & 1252    & 16724           &  -     & -     \\
        MCV                   & 22403   & 26225           &  427   & 427   \\
        GLR                   & 250     & 23377           &  176   & 6490  \\
        CN-Celeb              & 1002    & 6307            &  -     & -     \\
		\hline \hline
		Total files           & \multicolumn{2}{c}{72633} &  \multicolumn{2}{c}{8089} \\
        Total size, Gb        & \multicolumn{2}{c}{22.5}  &  \multicolumn{2}{c}{2.4}  \\
        Total duration, hours & \multicolumn{2}{c}{209.5} &  \multicolumn{2}{c}{22.1} \\
        \hline \hline
	\end{tabular}
\end{table}

\subsection{Collection session}
Large replay parallel dataset is made of many recording sessions. One session consists of sampling a subset of data from the source datasets, selecting one playback device, multiple recording devices and pinning one environment.

\subsubsection{Session data}
For each session we randomly picked the subset of the source data, about 4-8 hours long. All selected files were merged into one big file that was played by a playback device. After finishing recording, we applied a \hyperref[sec:post-processing]{post-processing} step to cut the audio back and map the recorded utterances to their original metadata.

\subsubsection{Session devices}
We used different configurations of playback and recording devices to collect the database. 19 recording and 11 playback devices were used in total (see Table \ref{table:playback-statistic} and Table \ref{table:record-statistic}). For each session we picked one playback and multiple recording devices. The recording was carried out using standard iOS/Android audio recording API.

\subsubsection{Session environment}
\label{subsec:envs}
There are total 17 unique recording environments in LRPD dataset. They include 25 sessions from an anechoic room with a stochastic noise (caused by air conditioning system), and up to 8 sessions for the rest of 16 environments each. These 16 environments represent the rooms and apartments of different sizes, and we also provide 8 labels of known apartments in the dataset. For the data where it was impossible to recover the apartment id we just set the label \emph{unknown} and this is a mix of recordings from the rest 8 environments.  For all the environments recording distance varied from 8 to 100 cm and it was pinned for each session and was similar for every device in that session.



\subsection{Collection stand}
\begin{figure}[t]
	\centering
	\includegraphics[width=1.0\linewidth]{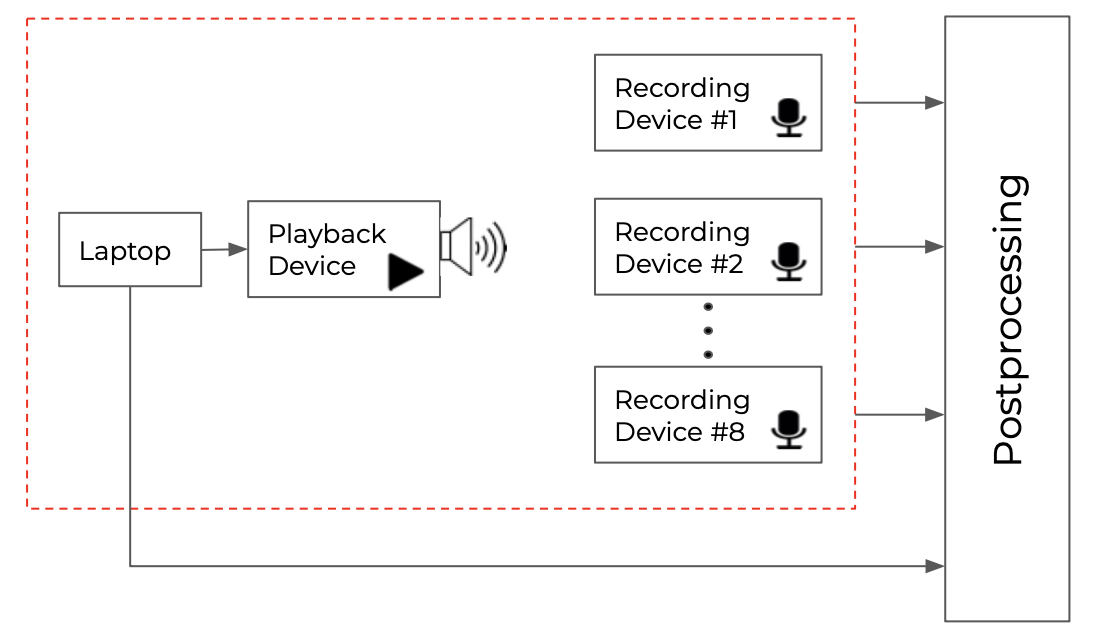}
	\caption{Schema of a stand}
    \label{fig:stand_schema}
\end{figure}

The collection stand was based on a laptop, Wi-Fi router, playback device and multiple recording devices. Playback device was connected to the laptop, and all the devices were connected to a power supply. Stand's schema is shown on Figure \ref{fig:stand_schema}.

\subsection{Post-processing dataset}
\label{sec:post-processing}

After collecting the data, we cut audio according to saved playback metadata. To double-check the quality of dataset we applied a cross-correlation function to the original file and its replayed copy.


\subsection{Dataset statistics}

Aggregated LRPD statistics and distributions across various playback devices (Table \ref{table:playback-statistic}), recording devices (Table \ref{table:record-statistic}), source data (Table \ref{table:source-statistic}) and recording environments are presented in corresponding tables.

\begin{table}[t]
	\centering
	\caption{Playback device statistic}
	\label{table:playback-statistic}
	\begin{tabular}{lcccc}
		\hline \hline
		& \multicolumn{3}{c}{\textbf{Ratio, \%}} \\
		\textbf{Playback} & \specialcell{trn aparts} & \specialcell{trn office} & \specialcell{val aparts} \\ 
		\hline
        srs xb12                   & 20.95        & 19.87      &   20.84  \\
        ginzzu gm 877b             & 18.85        & 9.05       &   18.82  \\
        lg pj2b                    & 14.53        & 24.42      &   14.64  \\
        jbl go3                    & 13.28        & 5.71       &   13.31  \\        
        jbl flip4                  & 12.26        & 11.92      &   12.37  \\        
        oklick ok-128              & 11.08        & 9.32       &   10.87  \\
        sharp                      & 6.70         & 0          &   6.59   \\
        defender enjoy s500        & 2.34         & 0          &   2.56   \\
        sps 609                    & 0            & 6.08       &   0      \\
        digma s16                  & 0            & 5.67       &   0      \\ 
        jbl clip3                  & 0            & 7.98       &   0      \\         
		\hline \hline
		Total files                & 469908      & 566832      &  195997  \\
		Total size, Gb             & 95.7        & 125.6       &  27.0    \\
		Total duration, hours      & 891.6       & 1170.6      &  251.8   \\
		\hline \hline
	\end{tabular}
\end{table}

\begin{table}[t]
	\centering
	\caption{Record device statistic}
	\label{table:record-statistic}
	\begin{tabular}{lcccc}
		\hline \hline
		& \multicolumn{3}{c}{\textbf{Ratio, \%}} \\
		\textbf{Record} & \specialcell{trn aparts} & \specialcell{trn office} & \specialcell{val aparts} \\ 
        \hline
        huawei matepad pro	       & 13.12	     &  0     	   &  13.14   \\
        huawei mate 40 pro	       & 13.11	     &  0          &  13.13   \\
        huawei mate 30 pro	       & 13.01	     &  0          &  13.01   \\
        huawei p smart z	       & 8.66	     &  4.59	   &  8.64    \\
        \hline
        honor 10x lite	           & 4.44	     &  4.56	   &  4.46    \\
        honor 30 pro+	           & 0           &  1.27	   &  0       \\
        \hline
        iphone 7	               & 4.19	     &  3.06	   &  4.24    \\
        iphone 8	               & 7.85	     &  0	       &  7.8     \\
        iphone 11 pro	           & 0	         &  9.18       &  0       \\
        iphone 11 pro max	       & 4.51	     &  9.18	   &  4.54    \\
        iphone 12 pro max	       & 0	         &  9.66       &  0       \\
        iphone xr	               & 7.72        &  8.47	   &  7.62    \\
        \hline
        samsung galaxy a01	       & 0           &  7.9	       &  0       \\
        samsung galaxy a51	       & 0	         &  4.85       &  0       \\
        samsung galaxy m21	       & 6.82	     &  3.12	   &  6.81    \\
        samsung galaxy s8+	       & 8.43	     &  10.37      &  8.4     \\
        samsung galaxy s20+	       & 3.9	     &  9.98	   &  3.94    \\
        \hline
        zte blade v2020	           & 4.23	     &  5.2	       &  4.26    \\
        sony xperia zx3            & 0	         &  8.6	       &  0       \\      
		\hline \hline
		Total files                & 469908      & 566832      &  195997  \\
		Total size, Gb             & 95.7        & 125.6       &  27.0    \\
		Total duration, hours      & 891.6       & 1170.6      &  251.8   \\
		\hline \hline
	\end{tabular}
\end{table}

\begin{table}[t]
	\centering
	\caption{Source dataset statistic}
	\label{table:source-statistic}
	\begin{tabular}{lcccc}
		\hline \hline
		& \multicolumn{3}{c}{\textbf{Ratio, \%}} \\
		\textbf{Source} & \specialcell{trn aparts} & \specialcell{trn office} & \specialcell{val aparts} \\ 
        \hline \hline  
        MCV                    &  36.3       &  29.4      &  16.68   \\
        GLR                    &  32         &  30.47     &  39.43   \\
        LibriSpeech            &  24.54      &  22.52     &  0       \\
        CN-Celeb               &  7.15       &  17.61     &  0       \\
        VCTK                   &  0          &  0         &  43.89   \\        
		\hline \hline
		Total files            & 469908      & 566832     &  195997  \\
		Total size, Gb         & 95.7        & 125.6      &  27.0    \\
		Total duration, hours  & 891.6       & 1170.6     &  251.8   \\
		\hline \hline
	\end{tabular}
\end{table}

\subsection{Dataset partitions}

We split LRPD into 3 parts: noised-train (\texttt{trn\_office}), clean-train (\texttt{trn\_aparts}) and test (\texttt{val\_aparts}). Noised train set contains files collected in an anechoic room. Clean train and test sets represent the rest 16 environments from Section \ref{subsec:envs}.


Metadata is currently included into dataset structure, and for information please check README file in dataset root.

\section{Experiments}


\subsection{Description}
We conducted experiments for two different tasks: 
\begin{enumerate}[label=Task \arabic*. , wide=0.5em,  leftmargin=*]
  \item Replay detection: binary classification of spoof / human. The goal of these experiments was to measure impact of adding new LRPD data for replay detection problem.
  \item Recording / Playback device classification (Device detection): two multi-label classification problems. Using the models obtained from device detector training we were able to explore embedding space using T-distributed Stochastic Neighbor Embedding (t-SNE) to visualize a distribution of 2-D embeddings on ASVspoof2017 eval subset.
\end{enumerate}

For more information on architectures, data setup, and training hyperparameters please refer to the baseline training pipeline.


\subsection{Datasets}


For the replay detection task, we used different combination of LRPD and ASVspoof 2017:

\begin{enumerate}
  \item{LRPD all train (office + aparts) + ASVspoof 2017 train}
  \item{LRPD all train (office + aparts)}
  \item{ASVspoof 2017 train}
\end{enumerate}


For the device detection task, we used LRPD all train (office + aparts) for training and LRPD eval for evaluation.

\subsection{Architecture}



We have chosen RawNet architecture \cite{jung2019rawnet} as a feature extractor model for both tasks, as we found it most suitable for detection of replay attacks. We have slightly changed initial RawNet architecture by replacing Gated recurrent unit (GRU) pooling with Statistical pooling layer and reducing model size using depth multiplier equals to 0.625. For each task we used different Fully Connected (FC) classifier head setup.

\subsection{Training setup}

For both tasks we used Adam optimizer with the following learning rate (lr) schedule: constant value of lr [1e\textsuperscript{-3}, 5e\textsuperscript{-4}, 1e\textsuperscript{-4}, 5e\textsuperscript{-5}] was dropped down after each 4 epochs, and we trained each model for 16 epochs in total. When training Task 1 models we sampled even number of utterances per class (replay/human) and even number of utterances from each dataset, forming batches of size 64. While for the Task 2 we sampled 4 replayed utterances with the same source utterance and cut them so they maintain aligned resulting in 4 x 32 batch size for Task 2.

Cross-entropy (CE) loss was used for both tasks, except that for Task 2 device detection we summed up CE losses from two simultaneous tasks: playback device classification and recording device classification:
\[ \mathcal{L}_{device-detection} = \mathcal{L}_{playback} + \mathcal{L}_{recording} \]

We augmented training data using noises from MUSAN \cite{snyder2015musan}, DCASE \cite{DCASE2017challenge} and DEMAND \cite {thiemann_joachim_2013_1227121} datasets. One random noise was added on-the-fly to each utterance with 0.5 probability and a random SNR uniformly sampled from 3-15 dB range.

\section{Results}

\subsection{Task 1. Replay detection}

The testing results of replay detector on the LRPD eval, ASVspoof17 eval and ASVspoof17 dev datasets are presented in Table \ref{table:detailed-results}. Adding LRPD data in training set may results in increasing of VAS system accuracy even on out-of-domain tests, such as ASVspoof17 eval, where EER drops from 13.94\% to 11.91\%.



\begin{table}[t]
	\centering
	\caption{Results on replay detection task.}
	\label{table:detailed-results}
	\begin{tabular}{lcccc}
		\hline \hline
		& \multicolumn{3}{c}{\textbf{EER, \%}} \\
		\textbf{Datasets used} & \specialcell{LRPD\\eval} & \specialcell{ASV17\\eval} & \specialcell{ASV17\\dev} \\ 
		\hline
        LRPD all train              & {\bf 0.16}   & 17.18        &  27.84        \\
        ASV17 trn              & 21.70        & 13.94        &  {\bf 17.54}  \\
        LRPD all train + ASV17 trn  & 0.28         & {\bf 11.91}  &  18.63        \\
		\hline \hline
	\end{tabular}
\end{table}

\begin{figure}[t]
	\centering
	\includegraphics[width=1.0\linewidth]{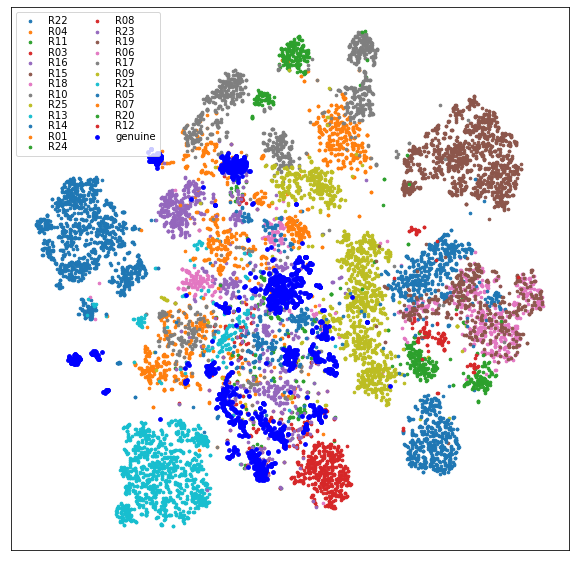}
	\caption{Device detector embedding space on ASVSpoof 2017 eval. R01-R25 stands for recording device ids in spoof utterances, and genuine class is bonafide.}
    \label{fig:embedding_space}
\end{figure}

\subsection{Task 2. Device detection}

t-SNE Visualisation of the device detector embeddings is presented on the Fig.\ref{fig:embedding_space}. Device detector was trained with LRPD train subset only, and visualized embeddings are extracted from out-of-domain ASVSpoof17 eval subset, for which recording and playback device meta information is given. We got 12.9\% EER on ASVSpoof17 eval subset (all with all protocol) for recording/playback pair classification task, using embeddings.

\section{Conclusions}
\label{sec:conclusions}

In this paper we presented the LRPD dataset, that was collected to advance future research on replay detection task. Compared with previous open-source datasets, the new corpus is larger, covers up-to-date recording and playback devices and contains more source data variety (speakers, languages). Using evaluations with the proposed data set, we found that the error of the baseline RawNet model drops by relative 15\% on target ASVSpoof2017 eval set, when trained with ASVSpoof2017 train and LRPD all train. We hope that proposed data will fuel further research in voice biometrics field by building more robust and protected systems.


\vfill\pagebreak

\bibliographystyle{IEEEbib}
\bibliography{AntispoofReplay.bib}

\end{document}